\documentclass[a4paper]{jpconf}
\usepackage{graphicx}
\usepackage{amssymb}
\usepackage{amsmath}
\usepackage{epsf}
\usepackage{bm}
\usepackage{cite}

\newtheorem{theorem}{Theorem}

\newtheorem{corollary}{Corollary}

\begin{document}

\title{Geometrization of Lie and Noether symmetries with applications in
Cosmology}
\author{Michael Tsamparlis}

\begin{abstract}
We derive the Lie and the Noether conditions for the equations of motion of
a dynamical system in a $n-$dimensional Riemannian space. We solve these
conditions in the sense that we express the symmetry generating vectors in
terms of the special projective and the homothetic vectors of the space.
Therefore the Lie and the Noether symmetries for these equations are
geometric symmetries or, equivalently, the geometry of the space is
modulating the motion of dynamical systems in that space. We give two
theorems which contain all the necessary conditions which allow one to
determine the Lie and the Noether symmetries of a specific dynamical system
in a given Riemannian space. We apply the theorems to various interesting
situations covering Newtonian 2d and 3d systems as well as dynamical systems
in cosmology.\newline
\newline
Keywords: Lie Symmetries, Noether symmetries, Cosmology\newline
Pacs - numbers:98.80.-k,04.20.-q, 45.20.D-,02.20Sv
\end{abstract}

\address{Faculty of Physics, Department of
Astronomy-Astrophysics-Mechanics,University of Athens, Panepistemiopolis, Athens 157 83, Greece}

\ead{mtsampa@phys.uoa.gr}

\section{Introduction}

In a Riemannian space the affinely parameterized geodesics are determined
uniquely by the metric. Therefore one should expect a close relation between
the geodesics as a set of homogeneous ordinary differential equations (ODE)
linear in the highest order term and quadratic non-linear in first order
terms, and the metric as a second order symmetric tensor. A system of such
ODEs is characterized ( not fully) by its Lie symmetries and correspondingly
a metric is characterized (again not fully) by its collineations. Therefore
it is reasonable to expect that the Lie symmetries of the system of geodesic
equation of a metric will be closely related with the collineations of the
metric. That such a relation exists it is easy to see by the following
simple example. Consider on the Euclidian plane a family of straight lines
parallel to the $x-$axis. These curves can be considered either as the
integral curves of the ODE\ $\frac{d^{2}y}{dx^{2}}=0$ or as the geodesics of
the Euclidian metric $dx^{2}+dy^{2}$. Subsequently consider a symmetry
operation defined by a reshuffling of these lines without preserving
necessarily their parametrization. According to the first interpretation
this symmetry operation is a Lie symmetry of the ODE\ $\frac{d^{2}y}{dx^{2}}%
=0$ and according to the second interpretation it is a (special) projective
symmetry of the Euclidian two dimensional metric.

What has been said for a Riemannian space can be generalized to a space in
which there is only a linear connection. In this case the geodesics are
called autoparallels (or paths) and they comprise again a system of ODEs
linear in the highest order term and quadratic non-linear in the first order
terms. In this case one is looking for relations between the Lie symmetries
of the autoparallels and the projective collineations of the connection.

A Lie point symmetry of an ordinary differential equation (ODE) is a point
transformation in the space of variables which preserves the set of
solutions of the ODE \cite{BlumanbookODES,OlverBook,StephanibookODES}. If we
look at these solutions as curves in the space of variables, then we may
equivalently consider a Lie point symmetry as a point transformation which
preserves the set of the solution curves. Applying this observation to the
geodesic curves in \ a Riemannian (affine) space, we infer that the Lie
point symmetries of the geodesic equations in any Riemannian space are the
automorphisms which preserve the set of these curves. However we know from
Differential Geometry that the point transformations of a Riemannian
(affine)\ space which preserve the set of geodesics are the projective
transformations. Therefore it is reasonable to expect a correspondence
between the Lie symmetries of the geodesic equations and the projective
algebra of the space.

The equation of geodesics in an arbitrary coordinate frame is a second order
ODE of the form%
\begin{equation}
\ddot{x}^{i}+\Gamma _{jk}^{i}\dot{x}^{j}\dot{x}^{k}+F(x^{i},\dot{x}^{j})=0
\label{de.0}
\end{equation}%
where $F(x^{i},\dot{x}^{j})$ is an arbitrary function of its argument and
the functions $\Gamma _{jk}^{i}$ are the connection coefficients of the
space. However equation (\ref{de.0}) is also the equation of motion of a
dynamical system moving in a Riemannian (affine) space under the action of a
velocity dependent force. According to the above argument we expect that the
Lie symmetries of the ODE (\ref{de.0}) for a given function $F(x^{i},\dot{x}%
^{j})$ will be a subalgebra of the projective algebra of the space. This
subalgebra is selected by means of certain constraint conditions which will
involve geometric quantities of the space and the function $F(x^{i},\dot{x}%
^{j})$ \cite%
{PrinceCrampin(1984)1,Aminova2006,Aminova2010,TsamparlisGRG,TsamparlisGRG2}.

The determination of the Lie point symmetries of a given system of ODEs
consists of two steps (a)\ the determination of the conditions which the
components of the Lie symmetry vector must satisfy and (b) the solution of
the system of these conditions. Step (a) is formal and it is outlined in
e.g. \cite{BlumanbookODES,OlverBook,StephanibookODES}. The second step is
the key one and, for example, in higher dimensions where one has a large
number of simultaneous equations the solution can be quite involved and
perhaps prohibitive. However if one expresses the system of Lie symmetry
conditions of (\ref{de.0}) in terms of collineation (i.e. symmetry)
conditions of the metric, then the determination of Lie symmetries it is
transferred to the geometric problem of determining the generators of a
collineation group of the metric and, as it will be shown below,
specifically the generators of the special projective group of the metric.
In this field there is a vast amount of work from Differential Geometry
waiting to be used. Indeed the projective symmetries are already known for
many spaces or they can be determined by existing general theorems. For
example the projective algebra and all its subalgebras are known for the
important case of spaces of constant curvature \cite{Barnes1993} and in
particular for the flat spaces. This implies that, for example, the Lie
symmetries of \emph{all} Newtonian dynamical systems as well as those of
Special Relativity can be determined by simple differentiation from the
known projective algebra of these spaces!

What has been said for the Lie point symmetries of (\ref{de.0}) applies also
to Noether point symmetries (provided (\ref{de.0}) follows from a
Lagrangian). The Noether symmetries are Lie point symmetries which satisfy
the additional constraint%
\begin{equation}
X^{\left[ 1\right] }L+L\frac{d\xi }{dt}=\frac{df}{dt}.  \label{L2p.3}
\end{equation}%
Noether symmetries form a closed subalgebra of the Lie point symmetries
algebra. In accordance to the above this implies that the Noether symmetries
will be related with a subalgebra of the projection algebra of the space
where `motion' occurs. As it will be shown this subalgebra is contained in
the homothetic algebra of the space. As it is well known with each Noether
point symmetry it is associated a conserved current (i.e. a Noether first
integral). This leads us to the importance conclusion that the (standard)
conserved quantities of a dynamical system depend on the space it moves and
the type of force $F(x^{i},\dot{x}^{j})$ which modulates the motion. In
particular in `free fall', that is when $F(x^{i},\dot{x}^{j})=0,$ the orbits
are the geodesics and the \emph{geometry} of the space is the sole factor
which determines the conserved quantities of motion. This conclusion is by
no means trivial and means that the space where motion occurs is not a
pathetic carrier of motion but it is the major modulator in the evolution of
a dynamical system. In other words there is a strong and deep relation
between Geometry of the space and Physics in that space!

The above matters have been discussed extensively in a series of interesting
papers by Aminova \cite{Aminova1994,Aminova1995,Aminova2000} who has given a
partial answer. Furthermore in a recent work \cite{FerozeMahomedQadir} the
authors have considered the KVs of the metric and their relation to the Lie
symmetries of the system of affinely parameterized geodesics in maximally
symmetric spaces of low dimension. In the same paper a conjecture is made,
which essentially says that the maximally symmetric spaces of non-vanishing
curvature do not admit further Lie symmetries.

The purpose of the present work is to present recent results concerning the
above approach. More specifically we do the following

a. We derive the general expression for the Lie symmetry conditions of (\ref%
{de.0})  assuming a general (i.e. not necessarily symmetric) connection in
terms of geometric quantities

b. For the case that the function $F$ depends only on the coordinates, i.e. $%
F(x^{i})$ we give two theorems which establish the exact relation between
the projective/homothetic algebra of the space and the Lie/Noether symmetry
algebra of (\ref{de.0}).

c. We consider applications in Newtonian Physics, General Relativity and in
Cosmology

The structure of the paper is as follows. In section \ref{The symmetry
conditions using Lie symmetry methods} we determine the conditions for Lie
symmetries in covariant form. We find that the major symmetry condition
relates the Lie symmetries with the special projective symmetries of the
connection. A\ similar result has been obtained by Prince and Crampin in
\cite{PrinceCrampin(1984)1} using the bundle formulation of second order
ODEs. In section \ref{The conservative system} we apply these conditions in
the case of Riemannian spaces. We solve the symmetry conditions and in
Theorem \ref{The general conservative system} we give the Lie symmetry
vectors in terms of the collineations of the metric. In section \ref{The Lie
symmetries of geodesic equations in an Einstein space} we apply Theorem \ref%
{The general conservative system} to the case of spaces of constant
curvature of dimension $n$ and compute the complete set of Lie symmetries of
the system of affinely parameterized geodesic equations in these spaces. We
distinguish the case of the flat space and the spaces of non-zero curvature
and prove the validity of the conjecture made in \cite{FerozeMahomedQadir}.
In the remaining sections we consider applications in General Relativity and
in Cosmology.

\section{Collineations of Riemannian spaces}

\label{Collineations of Riemannian spaces}

A\ collineation in a Riemannian space is a vector field $\mathbf{X}$ which
satisfies an equation of the form%
\begin{equation}
\mathcal{L}_{X}\mathbf{A=B}  \label{L2p.2}
\end{equation}%
where $\mathcal{L}_{X}$ denotes Lie derivative, $\mathbf{A}$ is a geometric
object (not necessarily a tensor)\ defined in terms of the metric and its
derivatives (e.g. connection, Ricci tensor, curvature tensor etc.) and $%
\mathbf{B}$ is an arbitrary tensor with the same tensor indices as $\mathbf{A%
}$. The collineations in a Riemannian space have been classified by Katzin
et al. \cite{Katzin}. In the following we use only certain collineations.

A conformal Killing vector (CKV)\ is defined by the relation
\begin{equation}
\mathcal{L}_{X}g_{ij}=2\psi \left( x^{k}\right) g_{ij}.
\end{equation}%
If $\psi =0,$ $\mathbf{X}\ $is called a Killing vector (KV), if $\psi $ is a
non-vanishing constant a homothetic vector (HV) and if $\psi _{;ij}=0,$ a
special conformal Killing vector (SCKV). A CKV is called proper if it is not
a KV, HV or a SCKV.

A projective collineation (PC) is defined by the equation
\begin{equation}
\mathcal{L}_{X}\Gamma _{jk}^{i}=2\phi _{(,j}\delta _{k)}^{i}.
\end{equation}%
If $\phi =0$ the PC\ is called an affine collineation (AC) and if $\phi
_{;ij}=0$ a special projective collineation (SPC). A\ proper PC\ is a PC\
which is not an AC, HV or KV or SPC. The PCs form a Lie algebra whose ACs,
HV and KVs form subalgebras. It has been shown that if a metric admits a
SCKV then also admits a SPC, a gradient HV and a gradient KV \cite{HallR}.
We summarize the above in the Table \ref{Table1}

\begin{table}[tbp]
\caption{The projective algebra of a Riemannian space}
\label{Table1}
\begin{center}
\begin{tabular}{lll}
\br \textbf{Collineation} & $\mathbf{A}$ & $\mathbf{B}$ \\
\mr Killing vector (KV) & $g_{ij}$ & $0$ \\
Homothetic vector (HV) & $g_{ij}$ & $\psi g_{ij}, \psi _{,i}=0$ \\
Conformal Killing vector (CKV) & $g_{ij}$ & $\psi g_{ij},\psi ,_{i}\neq 0$
\\
{Projective collineation (PC)} & $\Gamma _{jk}^{i}$ & $2\phi _{(,j}\delta
_{k)}^{i},$ $\phi ,_{i}\neq 0$ \\
Special Projective collineation (SPC) & $\Gamma_{jk}^{i}$ & $2\phi
_{(,j}\delta _{k)}^{i},$ $\phi ,_{i}\neq 0$ and $\phi ,_{jk}=0$ \\
\br &  &
\end{tabular}%
\end{center}
\end{table}

In the following we shall need the symmetry algebra of spaces of constant
curvature. In \cite{Barnes1993} it has been shown that the PCs of a space of
constant non-vanishing curvature consist of proper PCs and KVs only and if
the space is flat then the algebra of the PCs consists of KVs/HV/ACs and
SPCs. For convenience we summarize these results in Table \ref{Table2}.

\begin{table}[tbp]
\caption{The projective algebra of the Euclidian space}
\label{Table2}
\begin{center}
\begin{tabular}{lll}
\br \textbf{Collineation} & \textbf{Gradient} & \textbf{Non-gradient} \\
\mr Killing vectors (KV) & $\mathbf{S}_{I}=\delta _{I}^{i}\partial _{i}$ & $%
\mathbf{X}_{IJ}=\delta _{\lbrack I}^{j}\delta _{j]}^{i}x_{j}\partial _{i}$
\\
Homothetic vector (HV) & $\mathbf{H}=x^{i}\partial _{i}~$ &  \\
Affine Collineation (AC) & $\mathbf{A}_{II}=x_{I}\delta _{I}^{i}\partial_{i}~
$ & $\mathbf{A}_{IJ}=x_{J}\delta _{I}^{i}\partial_{i}~$ \\
Special Projective collineation (SPC) &  & $\mathbf{P}_{I}=S_{I}\mathbf{H}.~$
\\
\br {Where the indices $I,J=1,2,\ldots,n$.} &  &
\end{tabular}%
\end{center}
\end{table}

The Lie point symmetries of all Newtonian dynamical systems are amongst the
vectors in the above table. Also the Noether point symmetries of all
Newtonian dynamical systems (and more general all systems moving in a flat
space) follow from the elements of the first two rows of Table 2 (apart form
some differences in sign depending on the signature of the metric).

\section{The Lie point symmetry conditions in an affine space}

\label{The symmetry conditions using Lie symmetry methods}

We consider the system of ODEs:
\begin{equation}
\ddot{x}^{i}+\Gamma _{jk}^{i}\dot{x}^{j}\dot{x}^{k}+\sum%
\limits_{m=0}^{n}P_{j_{1}...j_{m}}^{i}\dot{x}^{j_{1}}\ldots \dot{x}^{j_{m}}=0
\label{de.1}
\end{equation}%
where $\Gamma _{jk}^{i}$ are the connection coefficients of the space and $%
P_{j_{1}...j_{m}}^{i}(t,x^{i})$ are smooth polynomials completely symmetric
in the lower indices and derive the Lie point symmetry conditions in
geometric form using the standard approach. Equation (\ref{de.1}) is quite
general and covers most of the standard cases autonomous and non autonomous
and in particular equation (\ref{de.0}). Furthermore because the $\Gamma
_{jk}^{i}$'s are not assumed to be symmetric, the results are valid in a
space with torsion. Obviously they hold in a Riemannian space where the
connection coefficients are given in terms of the Christofell symbols.

The detailed calculation can be found in \cite{TsamparlisGRG2} and we shall
not be repeat it here. In the following we summarize for convenience the
results.

The terms $\dot{x}^{j_{1}}\ldots \dot{x}^{j_{m}}$ for $m\leq 4$ give the
equations:
\begin{align}
L_{\eta }P^{i}+2\xi ,_{t}P^{i}+\xi P^{i},_{t}+\eta ^{i},_{tt}+\eta
^{j},_{t}P_{.j}^{i}& =0 \\
L_{\eta }P_{j}^{i}+\xi ,_{t}P_{j}^{i}+\xi P_{j}^{i},_{t}+\left( \xi
,_{k}\delta _{j}^{i}+2\xi ,_{j}\delta _{k}^{i}\right) P^{k}+2\eta
^{i},_{t|j}-\xi ,_{tt}\delta _{k}^{i}+2\eta ^{k},_{t}P_{.jk}^{i}& =0 \\
L_{\eta }P_{jk}^{i}+L_{\eta }\Gamma _{jk}^{i}+\left( \xi ,_{d}\delta
_{(k}^{i}+\xi ,_{(k}\delta _{|d|}^{i}\right) P_{.j)}^{d}+\xi
P_{.kj,t}^{i}-2\xi ,_{t(j}\delta _{k)}^{i}+3\eta ^{d},_{t}P_{.dkj}^{i}& =0 \\
L_{\eta }P_{.jkd}^{i}-\xi ,_{t}P_{.jkd}^{i}+\xi ,_{e}\delta
_{(k}^{i}P_{.dj)}^{e}+\xi P_{.jkd,t}^{i}+4\eta ^{e},_{t}P_{.jkde}^{i}-\xi
_{(,j|k}\delta _{d)}^{i}& =0~
\end{align}%
and the conditions due to the terms $\dot{x}^{j_{1}}\ldots \dot{x}^{j_{m}}$
for $m>4$ are given by the following general formula:
\begin{align}
& L_{\eta }P_{j_{1}...j_{m}}^{i}+P_{j_{1}...j_{m}~,t}^{i}\xi +\left(
2-m\right) \xi _{,t}P_{j_{1}...j_{m}}^{i}+  \notag \\
& +\xi _{,r}\left( 2-\left( m-1\right) \right) P_{j_{1}...j_{m-1}}^{i}\delta
_{j_{m}}^{r}+\left( m+1\right) P_{j_{1}...j_{m+1}}^{i}\eta
_{,t}^{j_{m+1}}+\xi _{,j}P_{j_{1}...j_{m-1}}^{j}\delta _{j_{m}}^{i}=0.
\end{align}

We note the appearance of the term $L_{\eta }\Gamma _{jk}^{i}$ in these
expressions.

Eqn (\ref{de.0}) is obtained for $m=0,$ $~P^{i}=F^{i}$ \ in which case the
Lie symmetry conditions read:
\begin{align}
L_{\eta }P^{i}+2\xi ,_{t}P^{i}+\xi P^{i},_{t}+\eta ^{i},_{tt}& =0
\label{de.13} \\
\left( \xi ,_{k}\delta _{j}^{i}+2\xi ,_{j}\delta _{k}^{i}\right) P^{k}+2\eta
^{i},_{t|j}-\xi ,_{tt}\delta _{k}^{i}& =0  \label{de.14} \\
L_{\eta }\Gamma _{jk}^{i}-2\xi ,_{t(j}\delta _{k)}^{i}& =0  \label{de.15} \\
\xi _{(,j|k}\delta _{d)}^{i}& =0.  \label{de.16}
\end{align}%
If $F^{i}=0$ we obtain the Lie symmetry conditions for the geodesic
equations (see \cite{TsamparlisGRG2},\cite{LiegeodesicsNonlinearDynamics2010}%
) .

\section{The autonomous dynamical system moving in a Riemannian space}

\label{The conservative system}

We `solve' the Lie symmetry conditions (\ref{de.13}) - (\ref{de.16}) for an
autonomous dynamical system in the sense that we express them in terms of
the collineations of the metric of the space where motion occurs.

Equation (\ref{de.16}) means that $\xi _{,j}$ is a gradient KV of $g_{ij}.$
This implies that the metric $g_{ij}$ is decomposable. Equation (\ref{de.15}%
) means that $\eta ^{i}$ is a projective collineation of the metric with
projective function $\xi _{,t}.$ The remaining two equations are constraint
conditions, which relate the components $\xi ,n^{i}$ of the Lie symmetry
vector with the vector $F^{i}(x^{j})$. Equation (\ref{de.13}) gives
\begin{equation}
\left( L_{\eta }g^{ij}\right) F_{j}+g^{ij}L_{\eta }F_{j}+2\xi
_{,t}g^{ij}F_{j}+\eta _{,tt}^{i}=0.  \label{de.21a}
\end{equation}%
This equation is an additional restriction for $\eta ^{i}$ because it
relates it directly to the symmetries of the metric. Finally equation (\ref%
{de.14}) gives%
\begin{equation}
-\delta _{j}^{i}\xi _{,tt}+\left( \xi _{,j}\delta _{k}^{i}+2\delta
_{j}^{i}\xi _{,k}\right) F^{k}+2\eta _{,tj}^{i}+2\Gamma _{jk}^{i}\eta
_{,t}^{k}=0.  \label{de.21d}
\end{equation}

We conclude that the Lie symmetry equations are equations (\ref{de.21a}) ,(%
\ref{de.21d}) where $\xi (t,x)$ is a gradient KV of the metric and $\eta
^{i}\left( t,x\right) $ is a special Projective collineation of the metric
with projective function $\xi _{,t}$. We state this result in theorem \ref%
{The general conservative system} \cite{TsamparlisGRG2}.

\begin{theorem}
\label{The general conservative system} The Lie point symmetries of the
system of equations of motion of an autonomous system under the action of
the force $F^{j}(x^{i})$ in a general Riemannian space with metric $g_{ij},$
namely%
\begin{equation}
\ddot{x}^{i}+\Gamma _{jk}^{i}\dot{x}^{j}\dot{x}^{k}=F^{i}  \label{PP.01}
\end{equation}%
are given in terms of the generators $Y^{i}$ of the special projective
algebra of the metric $g_{ij}.$
\end{theorem}

If the force $F^{i}$ is derivable from a potential $V(x^{i}),$ so that the
equations of motion follow from the standard Lagrangian
\begin{equation}
L\left( x^{j},\dot{x}^{j}\right) =\frac{1}{2}g_{ij}\dot{x}^{i}\dot{x}%
^{j}-V\left( x^{j}\right)  \label{NPC.02}
\end{equation}%
with Hamiltonian%
\begin{equation}
E=\frac{1}{2}g_{ij}\dot{x}^{i}\dot{x}^{j}+V\left( x^{j}\right)  \label{NPC.3}
\end{equation}%
it can be shown that the Noether conditions are
\begin{eqnarray}
V_{,k}\eta ^{k}+V\xi _{,t} &=&-f_{,t} \\
\eta _{,t}^{i}g_{ij}-\xi _{,j}V &=&f_{,j} \\
L_{\eta }g_{ij} &=&2\left( \frac{1}{2}\xi _{,t}\right) g_{ij} \\
\xi _{,k} &=&0.
\end{eqnarray}

Last equation implies $\xi =\xi \left( t\right) $ and reduces the system as
follows%
\begin{eqnarray}
L_{\eta }g_{ij} &=&2\left( \frac{1}{2}\xi _{,t}\right) g_{ij}  \label{NPC.4}
\\
V_{,k}\eta ^{k}+V\xi _{,t} &=&-f_{,t}  \label{PP.01.6} \\
\eta _{i,t} &=&f_{,i}.  \label{PP.01.7}
\end{eqnarray}

Equation (\ref{NPC.4}) implies that $\eta ^{i}$ is a conformal Killing
vector of the metric provided $\xi _{,t}\neq 0.$ Because $g_{ij}$\ is
independent of $t$\ and $\xi =\xi \left( t\right) $\ the $\eta ^{i}$\ must
be is a HV of the metric. This means that $\eta ^{i}\left( t,x\right)
=T\left( t\right) Y^{i}\left( x^{j}\right) $\ where $Y^{i}$\ is a HV. If $%
\xi _{,t}=0$ then $\eta ^{i}$ is a Killing vector of the metric. Equations (%
\ref{PP.01.6}), (\ref{PP.01.7}) are the constraint conditions, which the
Noether symmetry and the potential must satisfy for the former to be
admitted. The above lead to the following theorem \cite{TsamparlisGRG2}.

\begin{theorem}
\label{The Noether Theorem}The Noether point symmetries of the Lagrangian (%
\ref{NPC.02}) are generated from the homothetic algebra of the metric $%
g_{ij} $.
\end{theorem}

More specifically, concerning the Noether symmetries, we have the following.

All autonomous systems admit the Noether symmetry $\partial _{t}~$whose
Noether integral is the Hamiltonian $E$ given in equation (\ref{NPC.3}). For
the rest of the Noether symmetries we have the following cases

\textbf{Case I }\ Noether point symmetries generated by the homothetic
algebra.

The Noether symmetry vector and the Noether function $G\left( t,x^{k}\right)
$ are%
\begin{equation}
\mathbf{X}=2\psi _{Y}t\partial _{t}+Y^{i}\partial _{i}~,~G\left(
t,x^{k}\right) =pt  \label{NPC.03}
\end{equation}%
where $\psi _{Y}$ is the homothetic factor of $Y^{i}~$($\psi _{Y}=0$ for a
KV\ and $1$ for the HV) and $p$ is a constant, provided the potential
satisfies the condition%
\begin{equation}
\mathcal{L}_{Y}V+2\psi _{Y}V+p=0.  \label{NPC.04}
\end{equation}

\textbf{Case II} \ Noether point symmetries generated by the gradient
homothetic Lie algebra i.e. both KVs and the HV must be gradient. \

In this case the Noether symmetry vector and the Noether function are%
\begin{equation}
\mathbf{X}=2\psi _{Y}\int T\left( t\right) dt\partial _{t}+T\left( t\right)
H^{i}\partial _{i}~~,~G\left( t,x^{k}\right) =T_{,t}H\left( x^{k}\right)
~+p\int Tdt  \label{NPC.05}
\end{equation}%
where $H^{i}$ is the gradient HV or a gradient KV, the function $T(t)$ is
computed from the relation~$~T_{,tt}=mT~\ $where $~m$ is a constant and the
potential satisfies the condition
\begin{equation}
\mathcal{L}_{H}V+2\psi _{Y}V+mH+p=0.  \label{NPC.06}
\end{equation}

Concerning the Noether integrals we have the following result (not including
the Hamiltonian)

\begin{corollary}
\label{The Noether Integrals}The Noether integrals of Case I and Case II are
respectively
\begin{equation}
I_{C_{I}}=2\psi _{Y}tE-g_{ij}Y^{i}\dot{x}^{j}+pt  \label{NPC.07}
\end{equation}%
\begin{equation}
I_{C_{II}}=2\psi _{Y}\int T\left( t\right) dt~E-g_{ij}H^{,i}\dot{x}%
^{j}+T_{,t}H+p\int Tdt.  \label{NPC.08}
\end{equation}%
where $E$ is the Hamiltonian given in (\ref{NPC.3}).
\end{corollary}

We remark that theorems \ref{The general conservative system} and \ref{The
Noether Theorem} do not apply to generalized symmetries \cite{Sarlet,Kalotas}%
.

It would be of interest to examine if the above close relation of the Lie
and the Noether symmetries of the second order ODEs of the form (\ref{de.1})
with the collineations of the metric is possible to be carried over to some
types of second order partial differential equations (PDEs). To this
question it is not possible to give a global answer, due to the complexity
of the study and the great variety of PDEs. However for the case of the
generalized heat equation the problem has been solved (see \cite%
{Thefirstheatequationpaper1}).

\section{The Lie symmetries of geodesic equations in an Einstein space}

\label{The Lie symmetries of geodesic equations in an Einstein space}

Spaces of constant curvature are Einstein spaces whose curvature scalar is a
constant. In this section we generalize the results of the last section to
proper Einstein spaces in which $R$ is not a constant. Suppose $X^{a}$ is a
projective collineation with projection function $\phi(x^{a}),$ so that $%
\mathcal{L}_{X}\Gamma_{bc}^{a}=\phi_{,b}\delta_{c}^{a}+\phi_{,c}%
\delta_{b}^{a}.$ For a proper Einstein space $(R\neq0)$ we have $R_{ab}=%
\frac{R}{n}g_{ab}$ from which follows:%
\begin{equation}
\mathcal{L}_{X}g_{ab}=\frac{n(1-n)}{R}\phi_{;ab}-\mathcal{L}_{X}(\ln
R)g_{ab}.  \label{NPP.2}
\end{equation}

Using the contracted Bianchi identity $(R^{ij}-\frac{1}{2}Rg^{ij})_{;j}=0$
it follows that in an Einstein space of dimension $n>2$ the curvature scalar
$R=$constant and (\ref{NPP.2}) reduces to:%
\begin{equation*}
\mathcal{L}_{X}g_{ab}=\frac{n(1-n)}{R}\phi _{;ab}.
\end{equation*}%
It follows that if $X^{a}$ generates either an affine or a special
projective collineation then $\phi _{;ab}=0$ hence $X^{a}$ reduces to a KV.
This means that proper Einstein spaces do not admit homothetic vector,
affine collineations or special projective collineations.

The above results and Theorem \ref{The general conservative system} lead to
the following conclusion:

\begin{theorem}
\label{Theorem on Einstein spaces}The Lie point symmetries of the geodesic
equations in a proper Einstein space of curvature scalar $R$ $(\neq 0)$ are
given by the vectors $\mathbf{X}=\mathbb{\xi }\mathbf{+}\mathbb{\eta }%
\mathbf{=}\xi (t,x)\partial _{t}+\eta ^{i}(t,x)\partial _{x^{i}}$ as follows:%
\newline
a. The function:%
\begin{eqnarray}
\xi (t,x) &=&Kt+L \\
\eta ^{i}(t,x) &=&D^{i}(x)
\end{eqnarray}%
where $D^{i}(x)$ is a KV\ of the metric \ and $K,L$
\end{theorem}

Theorem \ref{Theorem on Einstein spaces} extends and amends the conjecture
of \cite{FerozeMahomedQadir} to the more general case of Einstein spaces.

\section{Applications in Newtonian Physics}

One important problem to consider is to find all 2d and 3d dimensional
autonomous Newtonian dynamical systems, that is, all dynamical systems
moving in flat Euclidian space with equation of motion $\ddot{x}%
^{i}=F^{i}(x^{j})$ where $i,j=1,2$ or $1,2,3$ which admit Lie and Noether
symmetries and of course to determine the admitted symmetries. The answer to
this question consists of finding all the forces $F^{i}(x^{j})$ which result
in this property.

The importance of this problem is twofold. Indeed the knowledge of the Lie
symmetries makes possible the computation of the invariants of the dynamical
system; also the knowledge of the Noether symmetries gives the Noether
integrals (conserved currents)\ of the system. Both these data can be used
to simplify the equations of motion and even lead to their analytic solution.

The two dimensional case has been considered originally by \cite{Sen} and
the 3d case by \cite{DamianouSophocleous1999,Damianou2004}. Both cases have
been considered anew following the geometric approach of the previous
paragraphs. More specifically the 2d case has been considered in \cite%
{TsamparlisPaliathanasis2011J.Phys.A} and the 3d case in \cite{TPL2012}. In
both cases it has been shown that the existing results were incomplete.

It is to be noted that from the applications point of view the 2d case is
important because it applies to the mini super space which is particularly
useful in cosmology. Indeed in the standard cosmological model and much of
its extensions the resulting dynamical system reduces to a two dimensional
dynamical system in a flat 2d Lorentz space (see \cite%
{VasilakosTsamparlisPaliathanasis2011}). For convenience we collect the
results of the 3d case in Tables \ref{Table3} and \ref{Table4}. Concerning
the determination of all 3d Newtonian dynamical systems, which admit Noether
point symmetries and subsequently the ones which are integrable via Noether
integrals we have the results of the Table \ref{T5} and Table \ref{T6}.

\begin{table}[tbp]
\caption{First family of 3d Newtonian dynamical systems admitting Lie
symmetries}
\label{Table3}
\begin{center}
\begin{tabular}{ll}
\br {\small Lie symmetry} & $\mathbf{F}\left( x_{\mu },x_{\nu },x_{\sigma
}\right) $ \\
\mr $\frac{d}{2}t\partial _{t}+\partial _{\mu }$ & $e^{-dx_{\mu }}{\small f}%
_{\mu ,\nu ,\sigma }\left( x_{\nu },x_{\sigma }\right) $ \\
$\frac{d}{2}t\partial _{t}+\partial _{\theta _{\left( \mu \nu \right) }}$ & $%
e^{-d\theta _{\left( \mu \nu \right) }}{\small f}_{\mu ,\nu ,\sigma }\left(
r_{\left( \mu \nu \right) },x_{\sigma }\right) $ \\
$\frac{d}{2}t\partial _{t}+R\partial _{R}$ & $x_{\mu }^{1-d}{\small f}_{\mu
,\nu ,\sigma }\left( \frac{x_{\nu }}{x_{\mu }},\frac{x_{\sigma }}{x_{\mu }}%
\right) $ \\
$\frac{d}{2}t\partial _{t}+x_{\mu }\partial _{\mu }$ & $x_{\mu }^{1-d}%
{\small f}_{\mu ,\nu ,\sigma }\left( x_{\nu },x_{\sigma }\right) $ \\
$\frac{d}{2}t\partial _{t}+x_{\nu }\partial _{\mu }$ & $e^{-d\frac{x_{\mu }}{%
x_{\nu }}}\left[ \frac{x_{\mu }}{x_{\nu }}f_{\nu }\left( x_{\nu },x_{\sigma
}\right) +f_{\mu }\left( x_{\nu },x_{\sigma }\right) \right] \partial _{\mu
}+f_{\nu }\partial _{\nu }+f_{\sigma }\partial _{\sigma }$ \\
\br &
\end{tabular}%
\end{center}
\end{table}

\begin{table}[tbp]
\caption{Second family of 3d Newtonian dynamical systems admitting Lie
symmetries}
\label{Table4}
\begin{center}
\begin{tabular}{ll}
\br {Lie symmetry} & ${F}_{\mu }\left( x_{\mu },x_{\nu },x_{\sigma }\right) $
\\
\mr ${\small t\partial }_{\mu }$ & ${\small f}_{\mu ,\nu ,\sigma }\left(
x_{\nu },x_{\sigma }\right) $ \\
${\small t}^{2}{\small \partial }_{t}{\small +tR\partial }_{R}$ & $\frac{1}{%
x_{\mu }^{3}}{\small f}_{\mu ,\nu ,\sigma }\left( \frac{x_{\nu }}{x_{\mu }},%
\frac{x_{\sigma }}{x_{\mu }}\right) $ \\
${\small e}^{\pm t\sqrt{m}}{\small \partial }_{\mu }$ & ${\small -mx}_{\mu }%
{\small +f}_{\mu ,\nu ,\sigma }\left( x_{\nu },x_{\sigma }\right) $ \\
$\frac{1}{\sqrt{m}}{\small e}^{\pm t\sqrt{m}}{\small \partial }_{t}{\small +e%
}^{\pm t\sqrt{m}}{\small R\partial }_{R}$ & ${\small -}\frac{m}{4}{\small x}%
_{\mu }{\small +}\frac{1}{x_{\mu }^{3}}{\small f}_{\mu ,\nu ,\sigma }\left(
\frac{x_{\nu }}{x_{\mu }},\frac{x_{\sigma }}{x_{\mu }}\right) $ \\
\br &
\end{tabular}%
\end{center}
\end{table}

Noether symmetries\vspace{3mm}
\begin{table}[tbp]
\caption{First family of 3d Newtonian dynamical systems which admit Noether
symmetries}
\label{T5}
\begin{center}
\begin{tabular}{lll}
\br {Noether} & ${d=0}$ & ${\ d\neq 2}$ \\
\mr $\frac{d}{2}{\small t\partial }_{t}{\small +\partial }_{\mu }$ & $%
{\small c}_{1}{\small x}_{\mu }{\small +f}\left( x_{\nu },x_{\sigma }\right)
$ & ${\small e}^{-dx_{\mu }}{\small f}\left( x_{\nu },x_{\sigma }\right) $
\\
$\frac{d}{2}{\small t\partial }_{t}{\small +\partial }_{\theta _{\left( \mu
\nu \right) }}$ & ${\small \,c}_{1}{\small \theta }_{\left( \mu \nu \right) }%
{\small +f}\left( r_{\left( \mu \nu \right) },x_{\sigma }\right) $ & $%
{\small e}^{-d\theta _{\left( \mu \nu \right) }}{\small f}\left( r_{\left(
\mu \nu \right) },x_{\sigma }\right) $ \\
$\frac{d}{2}{\small t\partial }_{t}{\small +R\partial }_{R}$ & ${\small x}%
^{2}{\small f}\left( \frac{x_{\nu }}{x_{\mu }},\frac{x_{\sigma }}{x_{\mu }}%
\right) $ & ${\small x}^{2-d}{\small f}\left( \frac{x_{\nu }}{x_{\mu }},%
\frac{x_{\sigma }}{x_{\mu }}\right) $ \\
$\frac{d}{2}{\small t\partial }_{t}{\small +x}_{\mu }{\small \partial }_{\mu
}$ & ${\small c}_{1}{\small x}_{\mu }^{2}{\small +f}\left( x_{\nu
},x_{\sigma }\right) $ & ${\small \nexists }$ \\
$\frac{d}{2}{\small t\partial }_{t}{\small +x}_{\nu }{\small \partial }_{\mu
}$ & ${\small c}_{1}{\small x}_{\mu }{\small +c}_{2}\left( x_{\mu
}^{2}+x_{\nu }^{2}\right) {\small +f}\left( x_{\sigma }\right) $ & ${\small %
\nexists }$ \\
\br &  &
\end{tabular}%
\end{center}
\end{table}

\begin{table}[tbp]
\caption{Second family of 3d Newtonian dynamical systems which admit Noether
symmetries}
\label{T6}
\begin{center}
\begin{tabular}{lll}
\br Noether & $V\left( x,y,z\right) $ &  \\
\mr ${\small t\partial }_{\mu }$ & ${\small c}_{1}{\small x}_{\mu }{\small +f%
}\left( x_{\nu },x_{\sigma }\right) $ &  \\
${\small t}^{2}{\small \partial }_{t}{\small +tR\partial }_{R}$ & $\frac{1}{%
x_{\mu }^{2}}{\small f}\left( \frac{x_{\nu }}{x_{\mu }},\frac{x_{\sigma }}{%
x_{\mu }}\right) $ &  \\
${\small e}^{\pm t\sqrt{m}}{\small \partial }_{\mu }$ & ${\small -}\frac{m}{2%
}{\small x}_{\mu }^{2}{\small +c}_{1}{\small x}_{\mu }{\small +f}\left(
x_{\nu },x_{\sigma }\right) $ &  \\
$\frac{1}{\sqrt{m}}{\small e}^{\pm t\sqrt{m}}{\small \partial }_{t}{\small +e%
}^{\pm t\sqrt{m}}{\small R\partial }_{R}$ & ${\small -}\frac{m}{8}R^{2}%
{\small +}\frac{1}{x_{\mu }^{2}}{\small f}\left( \frac{x_{\nu }}{x_{\mu }},%
\frac{x_{\sigma }}{x_{\mu }}\right) $ &  \\
\br &  &
\end{tabular}%
\end{center}
\end{table}

In order to show the use of these tables are used we consider two
applications; one concerning the motion of a Newtonian system on a 2d space
and the second motion in a 3d space. Before we proceed we recall that in
order a 3d Newtonian dynamical system to be integrable via Noether point
symmetries it must admit at least 3 Noether first integrals.

\subsection{Newtonian motion on the two dimensional sphere}

The motion of a system moving on the two dimensional sphere under the
potential function $V\left( \theta ,\phi \right) $ has Lagrangian
\begin{equation}
L\left( \phi ,\theta ,\dot{\phi},\dot{\theta}\right) =\frac{1}{2}\left( \dot{%
\phi}^{2}+\mathrm{Sinn}^{2}\phi ~\dot{\theta}^{2}\right) -V\left( \theta
,\phi \right)  \label{MCC.01}
\end{equation}%
where
\begin{equation*}
\mathrm{Sinn}\phi =\left\{
\begin{array}{cc}
\mathrm{\sin }\phi & K=1 \\
\mathrm{\sinh }\phi & K=-1%
\end{array}%
\right. ~\mathrm{Cosn}\phi =\left\{
\begin{array}{cc}
\cos \phi & K=1 \\
\mathrm{\cosh }\phi & K=-1.%
\end{array}%
\right. ~
\end{equation*}%
The equations of motion are:%
\begin{eqnarray}
\ddot{\phi}-\mathrm{Sinn}\phi ~\mathrm{Cosn}\phi \mathrm{~}\dot{\theta}%
^{2}+V_{,\phi } &=&0 \\
\ddot{\theta}+2\frac{\mathrm{Cosn}\phi }{\mathrm{Sinn}\phi }~\dot{\theta}%
\dot{\phi}+\frac{1}{\mathrm{Sinn}^{2}\phi }V_{,\theta } &=&0.  \label{MCC.03}
\end{eqnarray}%
The problem to find all potential functions for which the system is
integrable via Noether point symmetries \ is to find all potentials for
which the system admits at least one extra Noether symmetry, in addition to
the trivial one which is the energy (Hamiltonian)%
\begin{equation*}
E=\frac{1}{2}\left( \dot{\phi}^{2}+\mathrm{Sinn}^{2}\phi ~\dot{\theta}%
^{2}\right) +V\left( \theta ,\phi \right) .
\end{equation*}%
\qquad

\begin{table}[tbp]
\caption{The potentials $V\left( \protect\theta ,\protect\phi \right) $ for
which the dynamical system (\protect\ref{MCC.01}) is integrable via Noether
point symmetries}
\label{T7}
\begin{center}
\begin{tabular}{lll}
\br ${\ V}\left( \theta ,\phi \right) $ & {\ Noether Integral} &  \\
\mr ${\small F}\left( \cos \theta \mathrm{Sinn}\phi \right) $ & ${\small I}%
_{CK_{e,h}^{1}}$ &  \\
${\small F}\left( \sin \theta \mathrm{Sinn}\phi \right) $ & ${\small I}%
_{CK_{e,h}^{2}}$ &  \\
${\small F}\left( \phi \right) $ & ${\small I}_{CK_{e,h}^{3}}$ &  \\
${\small F}\left( \frac{1+\tan ^{2}\theta }{\mathrm{Sinn}^{2}\phi ~\left(
a-b\tan \theta \right) ^{2}}\right) $ & ${\small aI}_{CK_{e,h}^{1}}{\small %
+bI}_{CK_{e,h}^{2}}$ &  \\
${\small F}\left( a\cos \theta \mathrm{Sinn}\phi -K~b\mathrm{Cosn}\phi
\right) $ & ${\small aI}_{CK_{e,h}^{1}}{\small +bI}_{CK_{e,h}^{3}}$ &  \\
${\small F}\left( a\sin \theta \mathrm{Sinn}\phi -K~b\mathrm{Cosn}\phi
\right) $ & ${\small aI}_{CK_{e,h}^{2}}{\small +bI}_{CK_{e,h}^{3}}$ &  \\
${\small F}\left(
\begin{array}{c}
\left( a\cos \theta -b\sin \theta \right) \mathrm{Sinn}\phi + \\
-K~c\mathrm{Cosn}\phi%
\end{array}%
\right) $ & {\small \thinspace }$%
aI_{CK_{e,h}^{1}}+bI_{CK_{e,h}^{2}}+cI_{CK_{e,h}^{3}}$ &  \\
\br &  &
\end{tabular}%
\end{center}
\end{table}

In order to do that we consider the kinematic metric $g_{ij}=\frac{1}{2}%
\left( \dot{\phi}^{2}+\mathrm{Sinn}^{2}\phi ~\dot{\theta}^{2}\right) $ in
the configuration space, \ which is defined by the kinematic part of the
Lagrangian and apply theorem \ref{The Noether Theorem}.\newline
This has been done in \cite{TPL2012} and the answer is given in Table \ref%
{T7}. Where $I_{CK^{1,2,3}}$ are the Noether Integrals
\begin{eqnarray*}
I_{CK^{3}} &=&\dot{\theta}\mathrm{Sinn}^{2}\phi .~,~\ I_{CK^{1}}=\dot{\phi}%
\sin \theta +\dot{\theta}\cos \theta \mathrm{Sinn}\phi \mathrm{Cosn}\phi \\
I_{CK^{2}} &=&\dot{\phi}\cos \theta -\dot{\theta}\sin \theta \mathrm{Sinn}%
\phi \mathrm{Cosn}\phi
\end{eqnarray*}

\begin{corollary}
A dynamical system with Lagrangian (\ref{MCC.01})~has one, two or four
Noether point symmetries hence Noether integrals.

Proof: For the case of the free particle we have the maximum number of four
Noether symmetries (the rotation group $so(3)$ plus the $\partial _{t}$). In
the case the potential is not constant the Noether symmetries are produced
by the non-gradient KVs with Lie algebra~$\left[ X_{A},X_{B}\right]
=C_{AB}^{C}X_{C}~$where $C_{12}^{3}=C_{31}^{2}=~C_{23}^{1}=1$ for $%
\varepsilon =1~$and $\bar{C}_{21}^{3}=\bar{C}_{23}^{1}=\bar{C}_{31}^{2}=1$
for $\varepsilon =-1.~$Because the Noether point symmetries form a Lie
algebra and the Lie algebra of the KVs is semisimple the system will admit
either none, one or three Noether symmetries generated from the KVs. The
case of three is when $V\left( \theta ,\phi \right) =V_{0}$ that is the case
of geodesics, therefore the Noether point symmetries will be (including $%
\partial _{t}$) either one, two or four.
\end{corollary}

\subsection{Newtonian motion in 3d space}

We consider the dynamical system defined by the equations of motion%
\begin{equation*}
\ddot{x}^{\mu }={\small -}\frac{m}{4}{\small x}_{\mu }{\small +}\frac{1}{%
x_{\mu }^{3}}{\small f}_{\mu ,\nu ,\sigma }\left( \frac{x_{\nu }}{x_{\mu }},%
\frac{x_{\sigma }}{x_{\mu }}\right)
\end{equation*}

This system is the well known Ermakov system which has a long history in the
older and recent literature \cite%
{MoyoL,HaasGoedet2001,AthorneC,GovinderL,GovinderL2}.

From Table 6 we find without any calculations that the system admits the Lie
symmetry $\mathbf{X}=\frac{1}{\sqrt{m}}{\small e}^{\pm t\sqrt{m}}{\small %
\partial }_{t}{\small +e}^{\pm t\sqrt{m}}{\small R\partial }_{R}.$

\section{Applications in cosmology}

In this section we consider applications of Theorem \ref{The general
conservative system} and Theorem \ref{The Noether Theorem} to dynamical
systems occurring in Cosmology.

\subsection{ Lie and Noether symmetries of Bianchi class A homogeneous
cosmologies with a scalar field}

The Bianchi models in the ADM\ formalism are described by the metric%
\begin{equation}
ds^{2}=-N^{2}(t)dt^{2}+g_{\mu \nu }\omega ^{\mu }\otimes \omega ^{\nu }
\label{BCA.1}
\end{equation}%
where $N(t)$ is the lapse function and $\{\omega ^{a}\}$ is the canonical
basis 1-forms which satisfy the Lie algebra $d\omega ^{i}=C_{jk}^{i}\omega
^{j}\wedge \omega ^{k}$ where $C_{jk}^{i}$ are the structure constants of
the algebra.

The spatial metric $g_{\mu \nu }$ splits so that $g_{\mu \nu }=\exp
(2\lambda )\exp (-2\beta )_{\mu \nu }~$where $\exp (2\lambda )$ is the scale
factor of the universe and $\beta _{\mu \nu }$ is a $3\times 3$ symmetric,
traceless matrix, which can be written in a diagonal form with two
independent quantities, known as the anisotropy parameters $\beta _{+},\beta
_{-}$, as follows:%
\begin{equation}
\beta _{\mu \nu }=diag\left( \beta _{+},-\frac{1}{2}\beta _{+}+\frac{\sqrt{3}%
}{2}\beta _{-},-\frac{1}{2}\beta _{+}-\frac{\sqrt{3}}{2}\beta _{-}\right) .
\end{equation}

It has been shown that the dynamics of the class A Bianchi models with a
scalar field is described by the Lagrangian
\begin{equation}
L=e^{3\lambda }\left[ R^{\ast }+6\lambda -\frac{3}{2}(\dot{\beta}_{1}^{2}+%
\dot{\beta}_{2}^{2})-\dot{\phi}^{2}+V(\phi )\right]  \label{BCA.5}
\end{equation}
where $R^{\ast }$ is the Ricci scalar of the $3$ dimensional spatial
hypersurfaces given by the expression:%
\begin{align*}
R^{\ast }& =-\frac{1}{2}e^{-2\lambda }\left[ N_{1}^{2}e^{4\beta
_{1}}+e^{-2\beta _{1}}\left( N_{2}e^{\sqrt{3}\beta _{2}}-N_{3}e^{-\sqrt{3}%
\beta _{2}}\right) ^{2}-2N_{1}e^{\beta _{1}}\left( N_{2}e^{\sqrt{3}\beta
_{2}}-N_{3}e^{-\sqrt{3}\beta _{2}}\right) \right] \\
& +\frac{1}{2}N_{1}N_{2}N_{3}(1+N_{1}N_{2}N_{3}).
\end{align*}

The constants $N_{1},N_{2},$ and $N_{3}$ are the components of the
classification vector $n^{\mu }$ and $\beta _{1}=-\frac{1}{2}\beta _{+}+%
\frac{\sqrt{3}}{2}\beta _{-},$ $\beta _{2}=-\frac{1}{2}\beta _{+}-\frac{%
\sqrt{3}}{2}\beta _{-}$.\newline
It is important to note that the curvature scalar $R^{\ast }$ does not
depend on the derivatives of the anisotropy parameters $\beta _{+},$ $\beta
_{-}$ , equivalently of $\beta _{1},\beta _{2}.$

The Euler Lagrange equations due to the Lagrangian (\ref{BCA.5}) are:%
\begin{align*}
\ddot{\lambda}+\frac{3}{2}\dot{\lambda}^{2}+\frac{3}{8}(\dot{\beta}_{1}^{2}+%
\dot{\beta}_{2}^{2})+\frac{1}{4}\dot{\phi}^{2}-\frac{1}{12}e^{-3\lambda }%
\frac{\partial }{\partial \lambda }\left( e^{3\lambda }R^{\ast }\right) -%
\frac{1}{2}V(\phi )& =0 \\
\ddot{\beta}_{1}+3\dot{\lambda}\dot{\beta}_{1}+\frac{1}{3}\frac{\partial
R^{\ast }}{\partial \beta _{1}}& =0 \\
\ddot{\beta}_{2}+3\dot{\lambda}\dot{\beta}_{2}+\frac{1}{3}\frac{\partial
R^{\ast }}{\partial \beta _{2}}& =0 \\
\ddot{\phi}+3\dot{\phi}\dot{\lambda}+\frac{\partial V}{\partial \phi }& =0
\end{align*}%
where a dot over a symbol indicates derivative with respect to $t.$

We apply Theorem \ref{The general conservative system} and Theorem \ref{The
Noether Theorem} in order to compute the Lie and the Noether symmetries of
class A Bianchi models with a scalar field. Similar (but incomplete) works
on that topic can be found in \cite%
{CapozzieloMarmoRubanoScudellaro,CotsakisLeachPantazi,VakiliKhosvraviSepangi}%
.

We consider the four dimensional Riemannian space with coordinates $%
x^{i}=\left( \lambda ,\beta _{1},\beta _{2},\phi \right) $ and metric%
\begin{equation}
ds^{2}=e^{3\lambda }\left( 12d\lambda ^{2}-3d\beta _{1}^{2}-3d\beta
_{2}^{2}-2d\phi ^{2}\right) .  \label{BCA.10}
\end{equation}

It is easy to show that this metric is the conformally flat FRW spacetime
whose special projective algebra consists of the non gradient KVs
\begin{align*}
Y^{1}& =\partial _{\beta _{1}},~Y^{2}=\partial _{\beta _{2}},~Y^{3}=\partial
_{\phi },~Y^{4}=\beta _{2}\partial _{\beta _{1}}-\beta _{1}\partial _{\beta
_{2}} \\
~Y^{5}& =\phi \partial _{\beta _{1}}-\frac{3}{2}\beta _{1}\partial _{\phi
},~Y^{6}=\phi \partial _{\beta _{2}}-\frac{3}{2}\beta _{2}\partial _{\phi }
\end{align*}%
and the gradient HV~
\begin{equation*}
H^{i}=\frac{2}{3}\partial _{\lambda }~,~\psi =1.
\end{equation*}

The Lagrangian is written $L=T-U(x^{i})~$where $T=\frac{1}{2}g_{ij}\dot{x}%
^{i}\dot{x}^{i}$ is the geodesic Lagrangian, the potential function is%
\begin{equation}
U(x^{i})=-e^{3\lambda }\left( V\left( \phi \right) +R^{\ast }\right)
\label{BCA.11}
\end{equation}%
and we have used the fact that the curvature scalar does not depend on the
derivatives of the coordinates $\beta _{1},\beta _{2}.$

We apply Theorem \ref{The general conservative system} and Theorem \ref{The
Noether Theorem} to determine the Lie and the Noether symmetries of the
dynamical system with Lagrangian (\ref{BCA.5}) in the following cases:%
\newline
Case 1. Vacuum. In this case $\phi =$constant.\newline
Case 2. Zero potential $V\left( \phi \right) =0,$ $\dot{\phi}\neq 0$\newline
Case 3. Constant Potential $V\left( \phi \right) =$constant, $\dot{\phi}\neq
0$\newline
Case 4. Arbitrary Potential $V\left( \phi \right) ,~\dot{\phi}\neq 0.$

The detailed calculations can be found in \cite{TsamparlisGRG2}. Below we
give the results for Bianchi I in the Table \ref{T8}, for Bianchi II in the
Table \ref{T9}, for Bianchi VI and VII in the Table \ref{T10} and for
Bianchi VIII/IX spacetimes in the Table \ref{T11}.

\begin{table}[tbp]
\caption{Bianchi I spacetime}
\label{T8}
\begin{center}
\begin{tabular}{lll}
\br \textbf{Bianchi I} & Noether Sym. & Lie Sym. \\
\mr Vacuum & $\partial _{t},~Y^{1},~Y^{2},~Y^{4}$ & $\partial
_{t},~t\partial _{t},~Y^{1,2,4},~H^{i}$ \\
& $2t\partial _{t}+H^{i},~t^{2}\partial _{t}+tH^{i}$ & $t^{2}\partial
_{t}+tH^{i}$ \\
Zero Pot. & $\partial _{t},Y^{1-6}~,~2t\partial _{t}+H^{i}$ & $\partial
_{t},~t\partial _{t},~Y^{1-6}~$ \\
& $t^{2}\partial _{t}+tH^{i}$ & $H^{i},~t^{2}\partial _{t}+tH^{i}$ \\
Constant Pot. & $\partial _{t},~Y^{1-6}~$ & $\partial _{t},~Y^{1-6},H^{i}$
\\
& $\frac{1}{C}e^{\pm Ct}\partial _{t}\pm e^{\pm Ct}H^{i}$ & $\frac{1}{C}%
e^{\pm Ct}\partial _{t}\pm e^{\pm Ct}H^{i}$ \\
Arbitrary Pot. & $\partial _{t},~Y^{1,2,4}$ & $\partial
_{t},~Y^{1,2,4}~,H^{i}$ \\
Exponential Pot. & $\partial _{t},~Y^{1,2,4}$ & $\partial
_{t},~Y^{1,2,4}~,H^{i}$ \\
& $2t\partial _{t}+H^{i}+\frac{4}{d}Y^{3}$ & $t\partial _{t}+\frac{2}{d}%
Y^{3} $ \\
\br &  &
\end{tabular}%
\end{center}
\end{table}

\begin{table}[tbp]
\caption{Bianchi II spacetime}
\label{T9}
\begin{center}
\begin{tabular}{lll}
\br\textbf{Bianchi II} & Noether Sym. & Lie Sym. \\
\mr Vacuum & $\partial _{t},~Y^{2}$ & $\partial _{t},~Y^{2}~$ \\
& $6t\partial _{t}+3H^{i}-5Y^{1}$ & $~\frac{1}{3}t\partial
_{t}+H^{i},~t\partial _{t}-Y^{1}$ \\
Zero Pot. & $\partial _{t},~Y^{2},~Y^{3},~Y^{6}~$ & $\partial
_{t},~Y^{2},~~Y^{3},~Y^{6}$ \\
& $6t\partial _{t}+3H^{i}-5Y^{1}$ & $\frac{1}{3}t\partial
_{t}+H^{i},~t\partial _{t}-Y^{1}$ \\
Constant Pot. & $\partial _{t},~Y^{2},~Y^{3},~Y^{6}$ & $\partial
_{t},~Y^{2},~Y^{3},~Y^{6}$ \\
&  & $3H^{i}+Y^{1}$ \\
Arbitrary Pot. & $\partial _{t},~Y^{2}$ & $\partial
_{t},~Y^{2},~3H^{i}+Y^{1} $ \\
Exponential Pot. & $\partial _{t},~Y^{2}$ & $\partial
_{t},~Y^{2},~3H^{i}+Y^{1}$ \\
& $2t\partial _{t}+H^{i}-\frac{5}{3}Y^{1}+\frac{4}{d}Y^{3}$ & $t\partial
_{t}+\frac{2}{d}Y^{3}$ \\
\br &  &
\end{tabular}%
\end{center}
\end{table}

\begin{table}[tbp]
\caption{Bianchi VI/VII spacetimes}
\label{T10}
\begin{center}
\begin{tabular}{lll}
\br \textbf{Bianchi VI}$_{0}~/$\textbf{\ VII}$_{0}$ & Noether Sym. & Lie Sym.
\\
\mr Vacuum & $\partial _{t},~~6t\partial _{t}+3H^{i}+$ & $\partial
_{t},~H^{i}+\frac{1}{3}Y^{1}+\frac{\sqrt{3}}{3}Y^{2}$ \\
& $~-2Y^{1}-2\sqrt{3}Y^{2}$ & $2t\partial _{t}-Y^{1}-\sqrt{3}Y^{2}$ \\
Zero Pot. & $\partial _{t},~Y^{3}~,6t\partial _{t}+3H^{i}+$ & $\partial
_{t},~H^{i}+\frac{1}{3}Y^{1}+\frac{\sqrt{3}}{3}Y^{2}$ \\
& $~-2Y^{1}-2\sqrt{3}Y^{2}$ & $~Y^{3},2t\partial _{t}-Y^{1}-\sqrt{3}Y^{2}$
\\
Constant Pot. & $\partial _{t},~Y^{3}$ & $\partial _{t},~Y^{3},~H^{i}+\frac{1%
}{3}Y^{1}+\frac{\sqrt{3}}{3}Y^{2}$ \\
Arbitrary Pot. & $\partial _{t}~$ & $\partial _{t},~H^{i}+\frac{1}{3}Y^{1}+%
\frac{\sqrt{3}}{3}Y^{2}$ \\
Exponential Pot. & $\partial _{t}~,~6t\partial _{t}+3H^{i}-2Y^{1}+$ & $%
\partial _{t},~H^{i}+\frac{1}{3}Y^{1}+\frac{\sqrt{3}}{3}Y^{2}$ \\
& ~~$-2\sqrt{3}Y^{2}+\frac{6}{d}Y^{3}$ & $t\partial _{t}+\frac{1}{d}Y^{3}$
\\
\br &  &
\end{tabular}%
\end{center}
\end{table}

\begin{table}[tbp]
\caption{Bianchi VIII/IX spacetimes}
\label{T11}
\begin{center}
\begin{tabular}{lll}
\br \textbf{Bianchi VIII} & Noether Sym. & Lie Sym. \\
\br Vacuum & $\partial _{t}$ & $\partial _{t},~\frac{2}{3}t\partial
_{t}~+H^{i}$ \\
Zero Pot. & $\partial _{t},~Y^{3}$ & $\partial _{t},~Y^{3},~\frac{2}{3}%
t\partial _{t}~+H^{i}~$ \\
Constant Pot. & $\partial _{t},~Y^{3}$ & $\partial _{t},~Y^{3}$ \\
Arbitrary Pot. & $\partial _{t}$ & $\partial _{t}$ \\
\br \textbf{Bianchi IX} & Noether Sym. & Lie Sym. \\
\mr Vacuum & $\partial _{t}$ & $\partial _{t}$ \\
Zero Pot. & $\partial _{t},~Y^{3}$ & $\partial _{t},~Y^{3}$ \\
Constant Pot. & $\partial _{t},~Y^{3}$ & $\partial _{t},~Y^{3}$ \\
Arbitrary Pot. & $\partial _{t}$ & $\partial _{t}$ \\
\br &  &
\end{tabular}%
\end{center}
\end{table}

\section{Conclusion}

\label{Conclusions}

The Lie and the Noether symmetry vectors of the equations of motion in a
general Riemannian space are not independent dynamical quantities but
instead they are determined completely by the geometry of the space. Indeed
Theorem \ref{The general conservative system} and Theorem \ref{The Noether
Theorem} show that the Lie point symmetries are\ elements of the special
projective algebra of the space and the Noether point symmetries elements of
the homothetic algebra of the space. The selection of the particular vectors
for a given `force' modulating the motion of a particular system is done by
means of certain compatibility conditions of differential nature which can
easily be managed. Using the above theorems we have determined all 2d and 3d
Newtonian dynamical systems which admit Lie and Noether point symmetries and
have demonstrated how the results can be used to determine these symmetries
in specific situations. Furthermore we have applied the theorems in the case
of Cosmology and have determined the Lie and the Noether point symmetries of
the Bianchi Class A homogeneous models. More applications of these theorems
in cosmology can be found in \cite%
{VasilakosTsamparlisPaliathanasis2011,Vasilakosandus2011}, where one is able
to determine analytic solutions using the integration of field equations by
means of Noether integrals.

\ack This research was partially funded by the University of Athens Special
Account of Research Grants no 10812.

\end{document}